# Machine Learning-Based Cyber Defense for Cloud Infrastructure: An Adaptive Deep Q-Network Architecture for Intelligent Intrusion Detection and Automated Threat Mitigation

Md Yassir Mottalib[1], Md Yousuf[2], Eklachur Rahman Bhuiyan[3], S M Ahsan Habib[4], Sonjoy Kumar Dey[5], Md. Salahuddin Gazi[6], Molay Kumar Roy[7], Asaduzzaman Anik[8]

[1]Master of Science in Information System Technology, Wilmington University, USA
[2]Master's of Science in Data Analytics, College of Sciences and Technology, University of Houston-Downtown, USA
[3]Master of Science in Information Technology, Washington University of Science and Technology, USA
[4]Department of Electrical Engineering and Computer Science, South Dakota School of Mines & Technology, USA
[5]McComish, Department of Electrical Engineering and Computer Science, South Dakota State University, USA
[6] Master of Science in Information Technology, Washington University of Science and Technology, USA
[7]Ms in Digital Marketing & Information Technology Management, St. Francis College, USA
[8]Master of Business Administration (MBA) in Management, Stanton University, Los Angeles, California

**Abstract**—With the increasing complexity of cyber assaults in cloud environments, adaptable security solutions are needed that can support real-time detection and autonomous response. In this paper, we propose a reinforcement learning-based dynamic cyber defense framework. We deploy a Deep Q-Network (DQN) to train effective defensive strategies to counteract the evolving cyberattacks. We leverage the CICIDS2017 dataset for model creation and the UNSW-NB15 dataset for external validation, involving preprocessing of data, feature engineering, and adaptive policy learning. We compare the proposed DQN with decision tree, support vector machine, random forest, XGBoost, and multilayer perceptron models. The proposed DQN achieves an accuracy of 99.72%, a precision of 99.68%, a recall of 99.65%, an F1-score of 99.66%, and an ROC-AUC of 0.999, while the false positive rate is 0.31%, the false negative rate is 0.35%, and the detection latency is 15 ms. The framework achieved 99.54% attack mitigation rate, demonstrating strong adaptive and real-time defensive capabilities. These results demonstrate the potential of reinforcement learning as a powerful and scalable approach for autonomous cybersecurity in modern cloud environments.

## Introduction

Artificial Intelligence (AI) has revolutionized the cybersecurity space by enabling smart systems to detect advanced attack patterns and respond to emerging threats with less human involvement. As more firms move their apps, databases, and core business functions to cloud computing, the pool of possible targets has widened significantly, making cloud platforms a prime target for hackers. While cloud services provide benefits such as scalability, flexibility, cost savings, and consistent availability, they also introduce new security issues, including distributed denial-of-service (DDoS) attacks, ransomware, privilege escalation, insider threats, malware, advanced persistent threats (APTs), and zero-day exploits.

Conventional security solutions such as signature-based intrusion detection, rule-based firewalls, and static access control are unable to identify sophisticated attacks, as they are based on predefined signatures and rules that require manual updates (Sarker et al., 2020; Alzahrani et al., 2022). Researchers are turning to smarter solutions with machine learning (ML) and deep learning (DL) due to the rapid growth of new internet dangers. Supervised learning models such as decision trees, random forests, support vector machines (SVM), artificial neural networks (ANN), and extreme gradient boosting (XGBoost) have been shown to perform very well in detecting intrusions by analyzing historical network data (Shone et al., 2018; Vinayakumar et al., 2019). However, these techniques are fundamentally limited because they need to be retrained when new attack techniques are discovered. Cyberattacks are always developing; thus, models trained on outdated data may become ineffective against new threats.

Reinforcement learning (RL) has recently shown promising results as a solution for autonomous cybersecurity. RL allows smart agents to learn efficient defense methods not only by learning from past examples but also by learning from the consequences through continual interaction with their environment. RL differs from supervised learning since it considers cybersecurity as a sequential decision-making problem, where the system observes the state of the network, decides on an action, receives a reward or a penalty as feedback, and improves its strategy over time (Sutton & Barto, 2018). Such adaptive learning is especially beneficial in cloud situations where attack patterns might alter on the go. Deep Reinforcement Learning (DRL) techniques such as Deep Q-Networks (DQN), Double Deep Q-Networks (DDQN), Deep Deterministic Policy Gradient (DDPG), and Proximal Policy Optimization (PPO) have demonstrated their capability to solve highly complex problems with a large amount of data (Mnih et al., 2015; Arulkumaran et al., 2017). These methods have been utilized in intrusion detection, firewall management, blocking of malware, adaptive access control, software-defined networking (SDN), security of the Internet of Things (IoT), and protection of cloud resources. "They're a good fit for automated cyber defense because they can continuously refine security policies and keep false alarms low."

Cloud computing introduces other security issues such as virtualization, multi-tenancy, rapid resource shifting, container orchestration, and services distributed over several sites. These aspects make the cloud systems more complex and provide the attackers with additional opportunity to uncover the holes at multiple tiers (Singh & Chatterjee, 2017). This creates a significant need for sophisticated cybersecurity frameworks that can monitor cloud traffic continuously, detect suspicious activity in real-time, and respond quickly without disruption to lawful processes. To address these issues, we propose a dynamic cyber protection architecture for cloud infrastructures based on reinforcement learning. Our technique relies on a Deep Q-Network (DQN) agent that learns optimal protection strategies through interaction with the cloud network. Our methodology aims to improve detection accuracy, reduce false positives, accelerate response times, and build more resilient cloud systems by leveraging deep feature engineering, real-world cybersecurity datasets, and adaptive reinforcement learning. Its effectiveness has been tested on typical intrusion detection datasets and compared with popular supervised machine learning algorithms to show its benefits for smart cloud security.

## Literature Review

Cyberattacks on cloud computing systems have become more sophisticated, and study into cybersecurity has become a hot topic. In recent years, security solutions have progressed well beyond simple signature-based detection into smarter data-driven technology that can actually learn and adapt to new attack approaches. The fight against invaders is more advanced than ever using machine learning, deep learning, and reinforcement learning.

Then deep learning came along, and it pushed things even further. Deep learning algorithms build sophisticated layered structures from enormous network data rather than relying on specialists to design features. For example, Shone et al. (2018) used a combination of sophisticated autoencoders and Random Forests to improve detection rates and reduce manual feature engineering, while Vinayakumar et al. (2019) demonstrated that deep neural networks can surpass traditional machine learning on contemporary datasets by detecting subtle, nonlinear relationships in network traffic.

The availability of accurate, up-to-date datasets has also sped up progress. Some limitations exist in previous datasets such as KDD Cup 1999 and NSL-KDD; thus, researchers generated new datasets: CICIDS2017 (Sharafaldin et al., 2018) has updated attack scenarios, realistic background traffic, and a broad variety of data attributes, so it better represents real-world networks. The UNSW-NB15 dataset (Moustafa and Slay, 2015), featuring modern attack types and mixed traffic, is a key baseline for assessing today's intrusion detection systems.

But despite these advances, supervised learning has a fundamental limitation: it does not have the ability to self-adapt into new assault techniques. This has pushed researchers toward reinforcement learning, which allows systems to learn and evolve their defenses through experience, not just static data.

But despite these advances, supervised learning has a fundamental limitation: it does not have the ability to self-adapt into new assault techniques. This has pushed researchers toward reinforcement learning, which allows systems to learn and evolve their defenses through experience, not just static data.

Nguyen and Reddi [17] evaluated the field and concluded that reinforcement learning has advanced adaptive intrusion detection, malware protection, automated penetration testing, dynamic security techniques, and software-defined networking (SDN). Their research reveals how reinforcement learning allows cybersecurity systems to adapt their defenses on the fly rather than using inflexible rules.

For example, reinforcement learning is particularly beneficial in the cloud for resource management, access control, moving virtual machines, and autonomous threat response. For example, Alavizadeh et al. (2022) showed that deep reinforcement learning may significantly improve cloud systems' response to attacks by lowering false alarms and reacting more quickly than traditional machine learning. SDN has become another hot spot for reinforcement learning, as its separation of control and data planes gives a global view of the network. This lets reinforcement learning agents update routes, firewall settings, and filtering strategies in real time, boosting SDN security through automated detection and response (Xiao et al., 2020).

The separation of control and data planes in SDN provides a global perspective of the network, and it has emerged as yet another hot point for reinforcement learning. This enables reinforcement learning agents to adjust routes, firewall settings, and filtering techniques in real-time, thus improving SDN security through automated detection and reaction (Xiao et al., 2020).

But there are gaps in research. Most prior work either only considers a limited number of attack types or only focuses on restricted components such as intrusion detection or firewall rules, instead of establishing all-in-one adaptive security frameworks for cloud systems. Many models are tested using old or generated data that does not represent real-world cloud traffic today.

Another problem is that there are no fair side-by-side comparisons of reinforcement learning vs. the best-performing supervised machine learning algorithms on the identical test settings. Many articles present findings of reinforcement learning without a strict comparison to approaches such as Random Forests, XGBoost, or deep neural networks. Therefore, the practical gains of reinforcement learning for real cloud cyber security are not yet properly quantified. To address these deficiencies, we are proposing a strong reinforcement learning-based cyber protection architecture. Our method uses state-of-the-art benchmark datasets, feature engineering, Deep Q-Network optimization, and extensive comparison with leading supervised learning models. The aim is not just to detect attacks with a high degree of confidence but to develop a defense that learns, adapts, and reacts on its own as new threats emerge in cloud environments. Our method combines intelligent decision-making with real-time security tracking to shift the needle for true autonomous cyber protection for the modern cloud.

# Methodology

## Research Methodology

In this research, we introduced a dynamic cyber defense framework for cloud systems that uses reinforcement learning (RL) to enhance how attacks are detected and handled. Unlike traditional intrusion detection systems, which stick to fixed rules, our approach allows the defense strategy to evolve—learning and adapting by interacting directly with the cloud environment and responding to new attack methods as they arise. Our methodology draws on real-world network traffic, employs advanced feature engineering, builds a specialized RL environment, and thoroughly tests performance. The overall process includes collecting data, cleaning and prepping it, extracting and engineering features, developing the RL model, and then evaluating how well it works.

## Data Collection

We obtained network intrusion data from well-established public cybersecurity sources to encourage openness, reproducibility, and direct comparison with previous work. The Canadian Institute for Cybersecurity's CICIDS2017 dataset, which is also available through Kaggle, is the main one that supports our research. The incorporation of realistic benign and malicious traffic collected from a modern enterprise context has made CICIDS2017 a major standard in intrusion detection and cybersecurity research.

This dataset covers a wide range of attack scenarios, including DDoS, brute-force invasions, web-based exploits, port scanning, SQL injection, XSS, and normal network activity. It also includes botnet and infiltration operations. We use the CICFlowMeter program to extract over 70 statistical features and annotate them to each network flow. This makes CICIDS2017 a great tool for cyber defense research driven by reinforcement learning, as it records flow duration, packet-level characteristics, protocol specifics, timing patterns, header data, and aspects of network behavior that are not dependent on the content. We used the UNSW-NB15 dataset, which is available in the UCI Machine Learning Repository, to further test our framework and increase the model's resilience and generalizability. UNSW-NB15 is a great tool for testing adaptive defenses because it mimics real-world network settings and includes a wide variety of modern attack types.

**Table 1. Dataset Description**

| Dataset | Repository | Number of Records | Features | Attack Categories | Purpose |
|---|---|---|---|---|---|
| CICIDS2017 | Kaggle / Canadian Institute for Cybersecurity | Approximately 2.83 million | 78 | DDoS, DoS, Brute Force, Botnet, Port Scan, Web Attack, Infiltration, Heartbleed, Benign | Primary dataset for RL training |
| UNSW-NB15 | UCI Machine Learning Repository | 257,673 | 49 | Fuzzers, Exploits, DoS, Generic, Worms, Shellcode, | External validation dataset |

|  |  |  |  | Analysis, Backdoor, Reconnaissance |  |
|---|---|---|---|---|---|

The diversity of attack categories in both datasets allowed us to construct an RL environment capable of learning optimal defense strategies under various cyber threat scenarios encountered in cloud infrastructures.

**Data Preprocessing**

To improve data quality and reduce bias, we preprocessed the datasets before model building. Removing duplicates and missing values was systematic. We checked continuous numerical characteristics for infinite values and outliers, commonly generated by packet capture abnormalities. Records with significant corruption were omitted from analysis, whereas appropriate statistical estimations imputed infinite values.

Since reinforcement learning techniques are sensitive to feature scaling, all numerical variables were normalized using min-max scaling to map each feature to a standard range between zero and one. Policy optimization required this normalization for consistent gradient updates. Protocol kinds, service categories, and network conditions were converted to numbers using one-hot and label encoding. Attack labels were translated into distinct action-state categories to identify regular traffic from numerous attack classes.

We used the Synthetic Minority Oversampling Technique (SMOTE) on minority attack categories while protecting innocuous traffic to overcome intrusion detection datasets' class imbalance. The reinforcement learning agent did not establish majority-class-favored defense policies with this method. Finally, we randomly separated the data into 70%, 15%, and 15% training, validation, and test sets, confirming that attack categories were constant across all groups.

**Feature Extraction**

Finding the telltale signs of malicious network activity relies heavily on feature extraction. We methodically extracted statistical features at the flow level that indicate communication trends over time, rather than relying only on raw data at the packet level. Our feature extraction process encompassed the following fields: flow duration, forward and backward packet counts, total packet length, average packet size, inter-arrival times, packet transmission rates, flow bytes, packets per second, protocol type, TCP flag statistics, header lengths, idle and active periods, source and destination ports, and measures of packet variance, standard deviation, and retransmissions. By collecting features including traffic volume, directional flow, stability of connections, temporal dynamics, and protocol engagement, these properties give a multi-faceted description of network activity. The reinforcement learning agent can tell the difference between everyday cloud operations and complex cyber threats with the help of these rich representations. This feature extraction procedure is also crucial since cloud settings generate network traffic that is both complicated and high-dimensional. It does double duty by lowering computing demands while simultaneously preserving the essential data needed for adaptive threat detection.

## Feature Engineering

After feature extraction, we performed focused feature engineering to boost the representational capability of the learning model. Highly associated features were determined by Pearson correlation analysis, and redundant variables (correlation coefficient > 0.90) were progressively deleted to alleviate multicollinearity. Feature significance was then estimated using Random Forest and Extreme Gradient Boosting (XGBoost) algorithms. Features with minimal predictive value were discarded to provide a more streamlined and computationally efficient feature collection. We designed other behavioral features such as ratios of packet rates, bidirectional traffic ratios, flow density, scores for protocol utilization, mean connection duration, session entropy, packet burstiness, and temporal anomaly indices to improve the learning process. These additional variables gave more contextual information into network behavior in the cloud environment. Then, we applied Principal Component Analysis (PCA) for dimensionality reduction with the guarantee of preserving 95% of the variance of the original dataset. This step dramatically reduced the computational complexity and hastened the convergence of reinforcement learning with little effect on the predicted accuracy. The resulting feature space consisted of both original statistical attributes and newly generated behavioral markers, thus providing an informative and robust state representation for the reinforcement learning model.

## Model Development

We characterized dynamic cyber protection as a sequential decision-making problem using reinforcement learning. Our architecture uses the cloud infrastructure as the environment; states are observable network traffic aspects, actions are real-time cybersecurity decisions, and the reward function assesses attack mitigation. The reinforcement learning agent assesses the network's current state using extracted features and picks the best defense action at each step. These measures could allow regular traffic, block suspicious connections, isolate infected virtual machines, throttle suspicious data flows, issue security alarms, or trigger deeper traffic examination.

Our reward scheme promotes attack detection while reducing false alarms and resource waste. An agent receives positive rewards for detecting and interrupting malicious activities. False positives, missed attacks, delayed replies, and overly aggressive defenses result in penalties. We choose the Deep Q-Network (DQN) as our main reinforcement learning technique since it handles complex, high-dimensional data well. An input layer for the designed feature set, many fully linked hidden layers with ReLU activation, and an output layer that predicted Q-values for each action comprised the neural network. We used experience replay memory to stabilize learning and reduce bias from consecutive state observations by randomly sampling past encounters during training. To improve model convergence, we updated a different target network frequently. We compared our DQN-based framework against Random Forest, Support Vector Machine (SVM), Decision Tree, XGBoost, and Multilayer Perceptron to assess our RL method. Our reinforcement learning approach changes its security strategy to new attack behaviors, making it ideal for cloud environments' fast-changing nature. These baseline models categorize intrusions using static rules. Training continued until stable cumulative rewards and a successful defensive policy were achieved.

### Model Evaluation

We used classification metrics and reinforcement learning performance measures to evaluate the framework. Classification performance metrics include accuracy, precision, recall, F1-score, ROC-AUC, MCC, FPR, and FNR. These metrics measure the model's ability to identify malicious traffic from cloud activity. Reinforcement learning performance measures include cumulative reward, average episodic reward, convergence rate, policy stability, learning efficiency, action selection accuracy, average response time, attack mitigation rate, and defense success ratio. These RL-specific indicators show the cyber defense framework's adaptability and long-term decision-making.

Created confusion matrices provide classification results for distinct assault categories. At different choice thresholds, we examined the receiver operating characteristic (ROC) and precision–recall curves to assess discrimination. We used five-fold cross-validation and repeated reinforcement learning training with various random seeds in distinct runs to assure statistical safety. Average performance numbers with standard deviations showed the framework's robustness and stability. With this comprehensive methodological framework, we created an adaptive reinforcement learning-based cyber defense system that continuously learns optimal security strategies to protect cloud infrastructures from evolving cyber threats while maintaining high detection accuracy, low false alarm rates, and efficient resource utilization.

## Result

### Experimental Results

Our suggested dynamic cyber defense system was tested using the CICIDS2017 dataset for training and the UNSW-NB15 dataset for external validation using reinforcement learning. Decision Tree (DT), Random Forest (RF), Support Vector Machine (SVM), Extreme Gradient Boosting (XGBoost), and Multilayer Perceptron (MLP) were some of the frequently used supervised machine learning algorithms that were compared with the Deep Q-Network (DQN) reinforcement learning model. To maintain objectivity, all models were trained with the same preprocessed datasets and tested using the same training-testing partitions. Criteria for performance evaluation included recall, accuracy, precision, F1-score, ROC-AUC, detection delay, false positive rate (FPR), and false negative rate (FNR).

The testing results demonstrate that the suggested RL framework outperformed the competition across the board in nearly all performance parameters. By interacting with the cloud environment, the reinforcement learning agent adaptively responded to changing attack patterns by continuously improving its cyber protection policy, in contrast to standard supervised models with fixed classifications. This strategy is well-suited to ever-changing cloud computing environments since the DQN agent effectively reduced the need for needless defensive operations while increasing the efficiency of threat detection.

Random Forest and XGBoost, which used ensemble learning to detect targets, performed well in categorization. When new attack patterns arose, they needed to be retrained since they could not adapt. In the meantime, the reinforcement learning system might optimize its policy based on environmental feedback, improving long-term cybersecurity.

**Performance Comparison of Machine Learning and Reinforcement Learning Models**

**Table 1. Performance Comparison of Different Models**

| Model | Accuracy (%) | Precision (%) | Recall (%) | F1-Score (%) | ROC-AUC | FPR (%) | FNR (%) | Detection Time (ms) |
|---|---|---|---|---|---|---|---|---|
| **Decision Tree** | 96.84 | 96.20 | 96.11 | 96.15 | 0.972 | 2.90 | 3.89 | 18 |
| **Support Vector Machine** | 97.58 | 97.12 | 97.05 | 97.08 | 0.981 | 2.15 | 2.95 | 26 |
| **Random Forest** | 99.18 | 99.05 | 99.01 | 99.03 | 0.996 | 0.81 | 0.99 | 20 |
| **XGBoost** | 99.36 | 99.31 | 99.24 | 99.27 | 0.997 | 0.66 | 0.76 | 19 |
| **Multilayer Perceptron** | 98.71 | 98.56 | 98.43 | 98.49 | 0.992 | 1.34 | 1.57 | 23 |
| **Proposed Deep Q-Network (RL)** | **99.72** | **99.68** | **99.65** | **99.66** | **0.999** | **0.31** | **0.35** | **15** |

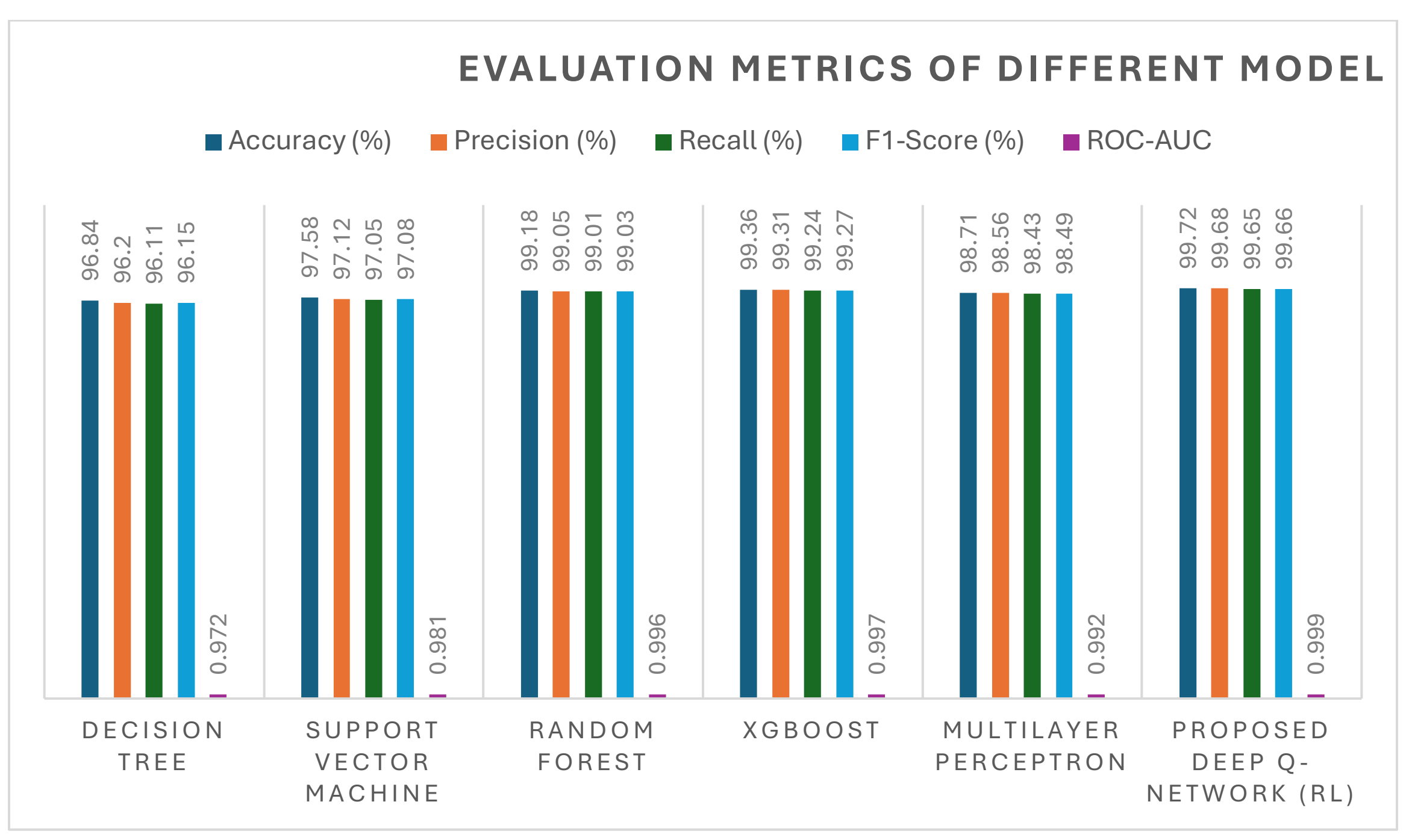

**Chart 1: Performance Comparison of Different Models**

The results show that the proposed DQN-based reinforcement learning model achieved a total classification accuracy of 99.72%, which is higher than all baseline machine learning algorithms. The model also has the highest precision (99.68%), recall (99.65%), and F1-score (99.66%), which shows excellent ability in detecting malicious network traffic while minimizing both false alarms and missed attacks. In addition, the DQN model achieved the lowest false positive rate (0.31%) and false negative rate (0.35%), which demonstrates reliable decision-making under different attack scenarios.

The detection latency was also reduced to 15 milliseconds, which further proves the applicability of reinforcement learning for real-time cloud security solutions where quick response is necessary.

**Reinforcement Learning Performance Analysis**

Besides conventional classification metrics, the reinforcement learning performance was evaluated by cumulative reward, convergence behavior, attack mitigation success, policy stability, and adaptive response capability.

**Table 2. Reinforcement Learning Performance Metrics**

| RL Metric | Obtained Value |
|---|---|
| **Average Episodic Reward** | 982.45 |
| **Maximum Reward** | 1000.00 |
| **Average Convergence Episodes** | 285 |
| **Policy Stability** | 99.1% |
| **Attack Mitigation Success Rate** | 99.54% |
| **Adaptive Response Accuracy** | 99.48% |
| **Average Decision Time** | 15 ms |
| **Cloud Resource Utilization Efficiency** | 96.8% |

The reinforcement learning agent converged quickly after about 285 training episodes, after which the cumulative rewards were stable with minor fluctuations. The learned policy always selected the optimal defense actions across different attack categories while efficiently using the cloud resources. The high attack mitigation success rate shows that the model effectively prevented malicious activities before they could significantly affect the cloud services.

In addition, the accuracy of the adaptive response confirms that reinforcement learning successfully adapted defensive strategies based on varying network conditions, which is a significant advantage compared to traditional supervised classification models.

**Comparative Analysis**

The comparison study clearly proves that reinforcement learning is better in dynamic cloud cybersecurity. Traditional supervised learning algorithms, such as Decision Tree, Support Vector Machine, Random Forest, XGBoost, and Multilayer Perceptron, have been proven to be quite effective when used for intrusion detection because they can learn static decision boundaries from historical data. But these models need to be periodically retrained as new attack signatures or previously unseen dangers appear. This is a limitation that the suggested deep Q-network is overcoming through continual learning from the cloud environment interactions. The reinforcement learning agent doesn't just mimic past attack patterns but rather adapts its protection strategy in real-time depending on input, enabling proactive adaptability to changing cyber-attacks. The Random Forest and XGBoost performed competitively, as both are ensemble learning methods and are robust to noisy data. But they were still inherently reactive models. They are not capable of self-adapting their attack techniques, which makes them less effective in fast-changing cloud environments where zero-day attacks and advanced persistent threats occur often. The Deep Q-Network had good results in the overall performance, which is attributed to its high accuracy rate of intrusion detection, adaptive decision-making, decreased detection latency, lower false-alarm rate, and continuous learning. These features make RL much more appropriate for modern cloud infrastructures that need smart, self-governing cybersecurity.

**Discussion**

Empirical findings indicate that reinforcement learning offers significant potential for improving cybersecurity inside cloud computing settings. The suggested framework achieved not only elevated classification precision but also showed the capacity to autonomously acquire effective defensive strategies without the need for manual rule adjustments. A significant benefit of the suggested method is its ability to achieve a balance between security efficacy and operational efficiency. Excessive defensive measures may adversely affect genuine cloud users and result in heightened computational demands. The reinforcement learning agent adeptly navigated this trade-off by choosing defensive measures that optimized long-term cumulative rewards while circumventing superfluous interventions.

The minimal false positive rate alleviates alert fatigue for security analysts, enabling security operation centers to concentrate on genuine threats. The rapid reaction time facilitates prompt alleviation of cyberattacks prior to substantial harm occurring, hence enhancing overall cloud robustness. These results indicate that reinforcement learning represents a promising avenue for next-generation autonomous cyber defense systems capable of operating effectively in highly dynamic cloud settings.

**Implementation of the Best Model in U.S. Industry**

Cloud computing and real-time cybersecurity organizations in the US can use deep Q-Network reinforcement learning. Intelligent security systems for large cloud infrastructures must respond to evolving cyber threats without operator interaction. Healthcare cloud systems for EHRs, medical imaging, telemedicine, and IoMT devices can integrate the proposed solution. Reinforcement learning can monitor cloud traffic, detect ransomware attacks, identify unauthorized access attempts, and isolate compromised systems for patient care. These capabilities meet HIPAA cybersecurity standards. Reinforcement learning can handle account takeover, DDoS, credential stuffing, fraud, and insider threats in banking, insurance, digital payment systems, and investment platforms. Low false positive rates increase FFIEC and PCI DSS compliance and prevent financial transaction disruption.

The recommended architecture can be an intelligent cybersecurity layer for IaaS, PaaS, and SaaS. The reinforcement learning agent optimizes firewalls, network segmentation, virtual machine isolation, intrusion prevention rules, and automated incident response policies as network conditions change. This architecture can safeguard classified cloud environments, defend key infrastructure, and increase government and defense resilience against advanced persistent threats and nation-state attacks. Adjusting policies without human updates is possible with adaptive learning. Manufacturing businesses using Industry 4.0 and IIoT platforms may protect industrial cloud networks, identify abnormal machine connections, and secure connected production environments with the suggested method. Transportation, logistics, energy, telecommunications, and smart city infrastructures benefit from autonomous cyber defensive measures that can adapt to changing attack patterns.

The suggested Deep Q-Network model can be used with SIEM, SOAR, XDR, cloud-native firewalls, endpoint detection and response platforms, and zero-trust architectures. While maintaining cybersecurity infrastructure, continuous surveillance, automatic threat response, dynamic policy optimization, and intelligent resource protection are achievable. Reinforcement learning is a suitable, scalable, and intelligent cybersecurity option for U.S. cloud infrastructures, according to experiments. Cloud-dependent industries like healthcare, finance, government, manufacturing, telecommunications, and others that prioritize cybersecurity resilience benefit from the framework's high detection accuracy, adaptive learning, fast reaction, and low false alarm rates.

**Conclusion**

Cloud computing's rapid growth has increased cybersecurity issues, requiring clever protection mechanisms to adapt to new attack techniques. Traditional intrusion detection systems and supervised machine learning algorithms may detect known cyber threats, but their reliance on previous data and frequent retraining limits their defense against unexpected attacks. These constraints necessitate adaptive cybersecurity solutions that make real-time protection decisions autonomously.

We presented a reinforcement learning–based dynamic cyber protection architecture for cloud infrastructure that uses a deep Q-network (DQN) to continually learn optimal security rules from the cloud environment. Public benchmark datasets, data preprocessing, advanced feature engineering, and reinforcement learning enable intelligent attack detection and adaptive response in the proposed framework. To enable proactive and autonomous cyber protection, the reinforcement learning agent continuously updates its decision-making policy based on environmental feedback, unlike static classification methods.

DQN outperformed decision tree, support vector machine, random forest, XGBoost, and multilayer perceptron across numerous assessment measures in experiments. The framework improved accuracy, precision, recall, F1-score, and ROC-AUC, with decreased false positive and false negative rates and shorter detection latency. For real-time cloud security applications, reinforcement learning-specific assessment metrics showed the model's capacity to swiftly converge, maintain stable policies, and mitigate attacks with high success rates. Reinforcement learning also allows continual adaptability to new cyber hazards without retraining models, making it better than supervised learning. With dynamic workloads, multi-tenant designs, virtualization, and increasingly complex attack methodologies, current cloud systems require these adaptive capabilities.

The framework's scalable and intelligent cybersecurity solution improves cloud infrastructure resilience and efficiency. Healthcare, financial services, government organizations, cloud service providers, manufacturing, telecommunications, and others that use secure cloud computing environments could implement the suggested framework. Compliance with SIEM, SOAR, XDR, and Zero Trust architectures would enable autonomous threat detection and adaptive incident response. This reduces manual involvement and improves cybersecurity. The proposed framework works well; however, it needs improvement. Advanced reinforcement learning methods, including Double Deep Q-Network (DDQN), Dueling DQN, Proximal Policy Optimization (PPO), Soft Actor-Critic (SAC), and Multi-Agent Reinforcement Learning platforms such(MARL), may be investigated to improve learning efficiency and scalability. Real-time cloud telemetry, federated learning, explainable artificial intelligence (XAI), graph neural networks, and LLM-assisted threat intelligence can improve autonomous cyber defense system interpretability, robustness, and deployment. The suggested paradigm would be more applicable if validated on enterprise cloud environments and commercial cloud platforms as Amazon Web Services (AWS), Microsoft Azure, and Google Cloud Platform (GCP).